\documentstyle[12pt]{article}
\begin{document}
\baselineskip=18pt
\begin{center}
{\Large Low-Lying Excited States and Low-Temperature Properties of}
{\Large an Alternating Spin-1 / Spin-{$\frac{1}{2}$} Chain : A DMRG study}
\end{center}
\centerline {\bf Swapan K. Pati$^{1}$, S. Ramasesha$^{1,3}$ and 
Diptiman Sen$^{2,3}$}

\vskip 0.5 truecm

\centerline{\it $^1$ Solid State and Structural Chemistry Unit,} 
\centerline{\it Indian Institute of Science, Bangalore 560012, India}
\centerline{\it $^2$ Centre for Theoretical Studies, Indian Institute 
of Science,}
\centerline{\it Bangalore 560012, India}
\centerline{\it $^3$ Jawaharlal Nehru Centre for Advanced Scientific 
Research,}
\centerline{\it Jakkur Campus, Bangalore 560064, India}

\begin{abstract}

We report spin wave and DMRG studies of the ground and low-lying excited 
states of uniform and dimerized alternating spin chains. The DMRG procedure 
is also employed to obtain low-temperature thermodynamic properties of the 
system. The ground state of a 2N spin system with spin-1 and spin-{$\frac{1}
{2}$} alternating from site to site and interacting via an antiferromagnetic 
exchange is found to be ferrimagnetic with total spin $s_G=N/2$ from 
both DMRG and spin wave analysis. Both the studies also show that there 
is a gapless excitation to a state with spin $s_G-1$ and a gapped excitation 
to a state with spin $s_G+1$. Surprisingly, the correlation length in the 
ground state is found to be very small from both the studies for this gapless 
system. For this very reason, we show that the ground state can be 
described by a variational {\it ansatz} of the product type. DMRG analysis 
shows that the chain is susceptible to a conditional spin-Peierls' 
instability. The DMRG studies of magnetization, magnetic susceptibility
($\chi$) and specific heat show strong magnetic-field dependence. The product 
$\chi T$ shows a minimum as a function of temperature($T$) at low-magnetic 
fields and the minimum vanishes at high-magnetic fields. This low-field
behaviour is in agreement with earlier experimental observations.
The specific
heat shows a maximum as a function of temperature and the height of the 
maximum increases sharply at high-magnetic fields. It is hoped that these 
studies will motivate experimental studies at high-magnetic fields.
\end{abstract}

\section {Introduction}

  The low-dimensional magnetic systems exhibit a wide 
variety of exotic physical phenomena. Many of these interesting 
and novel phenomena were  first predicted
from theoretical studies on spin systems in one-dimension
\cite{hal83,bon87,aff89}.
Fascinating amongst these have been the spin-Peierls 
instability and the Haldane conjecture. These predictions
motivated a number of experimental efforts towards synthesis
of low-dimensional magnetic systems. The experimental 
measurements\cite{buy86,ma92,CuGeO3} on some of these systems have 
provided support
for the Haldane conjecture and also for the 
existence of spin-Peierls instability in quasi-one-dimensional systems. 
While experimental studies\cite{buy86,ma92} of 
the spin-1 antiferromagnet $Ni(C_{2}H_{8}N_{2})_{2}
NO_{2}(ClO_{4})$ clearly show the existence of Haldane
gap, the measurement of the magnetic properties 
in a series of quasi-one-dimensional compounds confirm the presence of
spin-Peirels instability at low temperatures\cite{instab}. In recent years,
purely inorganic systems that show low-dimensional behaviour 
have also been synthesized. The compound $CuGeO_3$ has been shown
to be a quasi-one-dimensional system exhibiting spin-Peierls 
instability\cite{CuGeO3}. Another class of compounds that have 
become important
in recent years are the systems with spin ladders. The compound
that closely approximates a spin ladder is vanadyl pyrophosphate
with molecular formula $(VO)_{2}P_{2}O_{7}$. A spin ladder with
hole doping is predicted to support hole binding based on
exchange energy considerations as two holes occupying the 
same rung of the ladder would be energetically favoured,
if the exchange constant for the rung is larger than
that along the leg. These systems are hence expected to 
exhibit non-BCS type of superconductivity\cite{rice96}.

Yet another challenge in the area of molecular magnetism 
has been to synthesize a molecular ferromagnet. While
there have been many models for ferromagnetic exchange 
in molecular systems\cite{stein76}, the actual synthesis of molecular
ferromagnets has been relatively recent. The search for
molecular ferromagnet has led to the discovery of many
interesting molecular magnetic systems. 
In recent years, quasi-one-dimensional bimetallic molecular magnets, 
with each unit cell containing two spins of different 
spin value have been synthesized\cite{kahn1}. These systems contain 
two transition metal ions per unit cell and have the 
general formula $ACu(pbaOH)(H_{2}O)_{3}.2H_{2}O$ with 
$pbaOH$=2-hydroxo-1,3-propylenebis(oxamato) and $A$ = $Mn$, $Fe$, $Co$, 
$Ni$ and belong to the alternating or mixed spin chain
family\cite{kahn2}. These alternating spin compounds are seen to exhibit 
ferrimagnetic behaviour. The ferromagnetic or ferrimagnetic 
alignment of spins 
in molecular systems is usually difficult to achieve due to the diffused
nature of the molecular orbitals (MOs) which the unpaired electrons
occupy. The direct exchange integrals involving MOs are much
smaller than that found in transition metal ions. However, even with
the smaller exchange integral, we could have a ferro or ferri magnetic
spin alignment if  the molecules are so arranged as to yield a 
small intermolecular transfer integral. In such a situation,
the kinetic stabilization of the electrons is rather small 
and parallel alignment of the effective spin of the electrons 
on the molecules is favoured leading to a magnetic ground state\cite{sinha}

Thermodynamic properties of these alternating spin compounds show very
interesting behaviour\cite{kahn2,kahn3}. In very low magnetic fields, 
these systems show
one-dimensional {\it ferrimagnetic} behaviour. The ${\chi}T$ vs $T$  ($\chi$
is the magnetic susceptibility and $T$ is the temperature)
plots show a rounded minimum. 
As the temperature is increased, ${\chi}T$ decreases sharply, goes  
through a minimum before increasing gradually. The temperature
at which the minimum occurs differs from compound to compound.  
The behaviour of field induced magnetization with temperature is also 
quite interesting as the ground state of the system is a magnetic state. 
At moderate magnetic fields, with increase in temperature, 
the magnetization slowly increases, shows a broad peak and then decreases.
Such behaviour is conjectured to be due to irregular variation in the
spin multiplicities of the energy levels of the system. Theoretically,
there has been an earlier study of the alternating spin sytem
in three dimensions that focuses on the dependence of Curie temperature
on the anisotropy of the spins\cite{iwa}.

Motivation for the present study comes from the above 
experimental observations on the quasi-one-dimensional
alternating spin systems. Theoretical studies of spin chains so far, 
have however 
been concerned mainly with the antiferromagnetic spin systems 
with unique site spin, exceptions being dilute spin impurity problems.
In this study, we first analyse the ground and low-lying
excited states of the mixed or alternating antiferromagnetic
spin chain by employing a spin wave theory. The spin wave theory is, 
however, not as accurate as the recently developed
density matrix renormalization group (DMRG) method which has proved to be 
the best numerical tool for one-dimensional spin systems\cite{white1}. 
Ground 
state energy per site, the spin excitation gap and the two-spin 
correlation functions obtained from this method have been 
found to be accurate to several decimal places, in the case of 
the spin-{$\frac{1}{2}$} Heisenberg antiferromagnet which is amenable 
to exact study. 
In the DMRG method, spin parity symmetry can be used to characterize 
the spin states along with the  $s^{z}_{tot.}$ as the good quantum numbers
\cite{white2}.
There have been many interesting studies of spin-{$\frac{1}{2}$} and 
spin-1 chains in recent years, employing the DMRG technique\cite{dmrg}.
The spin-{$\frac{1}{2}$} and spin-1 dimerized Heisenberg chains with 
nearest neighbour and next nearest neighbour antiferromagnetic interactions 
have been recently studied by us by this new and powerful technique\cite{skp}.
There have also been some recent studies on a system with one or more
spin-{$\frac{1}{2}$} 
impurities embedded randomly in a spin-1 chain\cite{impuritydmrg}, 
and solution of spin models dynamically coupled to dispersionless 
phonons\cite{caron} by the DMRG method.

The spin wave analysis, in this paper, is followed by a report of our
results from extensive DMRG studies on the alternating spin 
system. The DMRG calculations have been carried out on chains/rings
with alternate spin-1 and spin-{$\frac{1}{2}$} sites. Studies on the ground 
state and low-lying excited states are reported in detail.  
Furthermore, by resorting to full diagonalization of the DMRG Hamiltonian
matrix in different total $M_s$ sectors, we have also obtained
the low-temperature thermodynamic properties of the alternating
spin chain with periodic boundary condition. The thermodynamic 
properties we discuss will include
low and high field magnetization, magnetic susceptibility
and specific heat at low temperatures. These properties are compared 
with experimental studies on bimetallic chains.

The paper is organized as follows. In the next section, we present 
properties of the ground and low-lying excited states obtained 
from spin wave analysis and DMRG calculations on long alternating spin chain
 with and without dimerization.
We also show that the short correlation length allows
the ground state to be approximated by a variational {\it ansatz}.
In the third section, we present the low-temperature thermodynamic properties
of the alternating spin chain. We end the paper with a summary of all
the results.

\section{Properties of the Ground and Low-Lying Excited states}

\subsection{Spin-Wave analysis}

We begin with the Hamiltonian for a chain with spins $s_1$ and $s_2$ on
alternating sites.
\begin{equation}
H=J \sum \limits_{n} [{\bf \hat S}_{1,n} \cdot
{\bf \hat S}_{2,n} +{\bf \hat S}_{2,n} \cdot {\bf \hat S}_{1,n+1}]
\end{equation}
 
\noindent where the total number of sites ( or bonds) is $2N$, and we use 
periodic boundary conditions in this section with ${\bf \hat S}_{1,N+1}
={\bf \hat S}_{1,1}$. 

The notation is illustrated in fig.1.  We assume that $s_{1} > 
s_{2}$, and will use spin wave theory\cite{ander52} to compute the leading
order corrections to the state shown in fig.1, where, 
 the $z$-component of the spin is $s_{1}$ for the spin-$s_{1}$ sites and 
 $-s_{2}$ for spin-$s_{2}$ sites. The Holstein-Primakoff transformations
take the form
\begin{eqnarray}
{\bf \hat S}^{z}_{1,n}&=&s_{1} -{\hat a}^{\dag}_{n}{\hat a}_{n} \nonumber \\
{\bf \hat S}^{+}_{1,n}&=&\Biggl(\sqrt{2s_{1} -{\hat a}^{\dag}_{n}{\hat a}_{n}}
\Biggr) 
{\hat a}_{n} \nonumber \\
{\bf \hat S}^{-}_{1,n}&=&{\hat a}^{\dag}_{n} \Biggl(\sqrt{2s_{1} - 
{\hat a}^{\dag}_{n}
{\hat a}_{n}}\Biggr) 
\end{eqnarray}
\noindent for the spin-$s_{1}$ sites, and 
\begin{eqnarray}
{\bf \hat S}^{z}_{2,n}&=&-s_{2} +{\hat b}^{\dag}_{n}{\hat b}_{n} \nonumber \\
{\bf \hat S}^{+}_{2,n}&=&{\hat b}^{\dag}_{n}\Biggl(\sqrt{2s_{2}-
{\hat b}^{\dag}_{n}
{\hat b}_{n}}\Biggr) \nonumber \\
{\bf \hat S}^{-}_{2,n}&=&\Biggl(\sqrt{2s_{2}-{\hat b}^{\dag}_{n}{\hat b}_{n}}
\Biggr){\hat b}_{n}
\end{eqnarray}
\noindent for the spin-$s_{2}$ sites. We then rewrite the Hamiltonian
in terms of the bosonic operators ${\hat a}_{n}$, ${\hat b}_{n}$, 
${\hat a}^{\dag}_{n}$ and
${\hat b}^{\dag}_{n}$, expand to quadratic order and Fourier transform to get
\begin{equation}
H= -2NJs_{1}s_{2}+ 2J \sum \limits_{k} [ s_{1}{\hat b}^{\dag}_{-k}{\hat b}_
{-k}
+ s_{2}{\hat a}^{\dag}_{k}{\hat a}_{k} + \sqrt{s_{1}s_{2}} \cos(k/2)
({\hat a}_{k}{\hat b}_{-k}+{\hat a}^{\dag}_{k}
{\hat b}^{\dag}_{-k})]
\end{equation}
This can be transformed using the Bogoliubov transformation, in the form 
\begin{eqnarray}
{\hat c}_{k}&=&{\hat a}_{k}\cosh \theta_{k}+ {\hat b}^{\dag}_{-k}\sinh 
 \theta_{k} \nonumber \\
 {\hat d}_{k}&=&{\hat b}_{-k}\cosh \theta_{k} + {\hat a}^{\dag}_{k}\sinh 
\theta_{k} \nonumber \\
\tanh (2 \theta_{k}) &=& \frac{2\sqrt{s_{1}s_{2}}}{ s_{1}+s_{2}} \cos(k/2)
\end{eqnarray}
\noindent We then get
\begin{equation}
H= -2NJs_{1}s_{2} + \sum \limits_{k} [ \omega_{1k}{\hat c}^{\dag}_{k}
{\hat c}_{k} 
+\omega_{2k} {\hat d}^{\dag}_{k}{\hat d}_{k}+ z_{k}]
\end{equation}
\noindent where the mode energies $\omega_{ik}$ and zero-point energy
$z_{k}$ are given by  
\begin{eqnarray}
\omega_{1k}&=& J(-s_{1}+s_{2})+ \omega_{k} \nonumber \\ 
\omega_{2k}&=& J(s_{1}-s_{2})+ \omega_{k} \nonumber \\
 z_{k} &=& -J(s_{1}+s_{2}) + \omega_{k} \nonumber \\
\omega_{k}&=& J\sqrt{(s_{1}-s_{2})^{2} + 4s_{1}s_{2} \sin^{2}(k/2)}
\end{eqnarray} 

The ground state $|\psi_{0}>$ is the state annihilated by all the
operators ${\hat c}_{k}$ and ${\hat d}_{k}$, because $\omega_{1k}$ and 
$\omega_{2k}$ are
 positive for all $k$. We see that the modes denoted by $\omega_{1k}$ are
gapless at $k=0$, where they have a ferromagnetic dispersion
$\omega_{1k} \sim k^{2}$. The modes $\omega_{2k}$ are gapped for all
$k$, with a minimum gap $\Delta = 2J(s_{1}-s_{2})$ at $k=0$.

The ground state energy per bond is given by 
\begin{equation}
{\epsilon}_{0} =\frac{E_{0}}{2N}= -Js_{1}s_{2} +\frac{1}{2} 
\int \limits_{0}^{\pi} \frac{dk}{\pi}[-J(s_{1}+s_{2})+\omega_{k}] 
\end{equation}
For $s_{1}=1$ and $s_{2}=\frac{1}{2}$, we get ${\epsilon}_{0}=-0.718J$. 
The spin wave ground state $|\psi_{0}>$ can also be shown to be
an eigenfunction of the 
${\bf \hat S}^{z}_{tot.}$ operator with the eigenvalue $N(s_{1}-s_{2})$.
The sublattice magnetizations are given by the expectation values of
${\bf \hat S}^{z}_{1,n}$ and ${\bf \hat S}^{z}_{2,n}$,
\begin{eqnarray}
<{\bf \hat S}^{z}_{1,n}> &=& (s_{1}+\frac{1}{2}) -\frac{1}{2} \int 
\limits_{0}^{\pi}
{\frac{dk}{\pi} \frac{(s_{1}+s_{2})J}{\omega_{k}}} \nonumber \\
<{\bf \hat S}^{z}_{2,n}> &=& s_{1}-s_{2}-<{\bf \hat S}^{z}_{1,n}>
\end{eqnarray}
For $s_{1}=1$, $s_{2}=\frac{1}{2}$, we find $<{\bf \hat S}^{z}_{1,n}> 
= 0.695$ and
 $<{\bf \hat S}^{z}_{2,n}> = -0.195$.
The operator 
\begin{equation}
{\bf \hat S}^{+}_{tot.} = \sum \limits_{n} [{\hat b}^{+}_{n}(\sqrt{2s_{2}})+
 {\hat a}_{n}
(\sqrt{2s_{1}})] = {\hat c}_{0}(\sqrt{2N(s_{1}-s_{2}}),
\end{equation}
\noindent to linear order in the bosonic operators.
Since this annihilates $|\psi_{0}>$, which is an eigenstate of 
${\bf \hat S}^{z}_{tot.}$, we conclude that $|\psi_{0}>$ has
$s_{tot.}=s^{z}_{tot.}=N(s_{1}-s_{2})$. Thus the ground state
is a ferrimagnet.

We can show that the $\omega_{1k}$ modes are created by acting on
$|\psi_{0}>$
with ${\bf \hat S}^{-}_{k}= \frac{1}{\sqrt{N}} \sum \limits_{n} 
({\bf \hat S}^{-}_{1,n}+ {\bf \hat S}^{-}_{2,n}) e^{ikn}$, 
where $k \ne 0$. The resultant states have $s^{z}_{tot.}=N(s_{1}-s_{2}) -1$
and are also annihilated by ${\bf \hat S}^{+}_{tot.}$. Therefore we
conclude that these states have $s_{tot.}=s^{z}_{tot.}=N(s_{1}-s_{2})-1$. 
Since
$\omega_{1k}$ is a gapless branch, we further conclude that the system
has gapless excitations to states with spin $N(s_{1}-s_{2})-1$.
Similarly, the $\omega_{2k}$ modes are created by acting on $|\psi_{0}>$
with ${\bf \hat S}^{+}_{k}=\frac{1}{\sqrt{N}} \sum \limits_{n}
({\bf \hat S}^{+}_{1,n}+ {\bf \hat S}^{+}_{2,n})e^{-ikn}$, 
where $k \ne 0$. By a similar argument, we can show that these states have
$s_{tot.}=s^{z}_{tot.}=N(s_{1}-s_{2})+1$. The branch $\omega_{2k}$
is separated
from the ground state by a gap  $\Delta = 2J(s_{1}-s_{2})$. Fig.2 shows
these two excitation branches for $s_{1}=1$ and $s_{2}=\frac{1}{2}$

Finally, we can calculate the two-spin correlation functions. There are
three kinds of functions one can consider, namely $<{\bf \hat S}_{1,0} \cdot 
{\bf \hat S}_{1,n}>$, $<{\bf \hat S}_{2,0} \cdot {\bf \hat S}_{1,n}>$ 
and $<{\bf \hat S}_{2,0} \cdot {\bf \hat S}_{2,n}>$. We will consider only
the first case as an illustrative example. Since the ground state has
long range order ( with $s_{tot.}=s^{z}_{tot.}=N(s_{1}-s_{2})$ ), 
we consider the subtracted correlation function 
\begin{eqnarray}
<{\bf \hat S}_{1,0} \cdot {\bf \hat S}_{1,n}> - <{\bf \hat S}_{1,0}>
\cdot <{\bf \hat S}_{1,n}> &=& s_{1} < \psi_{0} |{\hat a}_{0} 
{\hat a}^{\dag}_{n} 
+ {\hat a}^{\dag}_{0} {\hat a}_{n}| \psi_{0}> \nonumber \\
&=& Js_{1} \int \limits_{0}^{\pi} \frac{dk}{\pi}(s_{1}+s_{2}) \frac{\cos kn}
{\omega_{k}}
\end{eqnarray}
\noindent For $n \rightarrow \infty$, we can show that the leading behaviour
of this is given by an exponentially decaying factor $e^{-n/\xi}$, where
$\xi^{-1} =\ln(s_{1}/s_{2})$. For $s_{1}=1$, $s_{2}=\frac{1}{2}$, we get
$\xi=1.44$. This remarkably short correlation length agrees well with
numerical results as will be seen later. It may also be compared to the 
much larger values for the pure spin-1 antiferromagnet in which 
$\xi \approx 6$\cite{aklt,taka}
and the pure spin-{$\frac{1}{2}$} antiferromagnet in which $\xi=\infty$.

We can use spin wave theory to study a dimerized model described by the
Hamiltonian
\begin{equation}
H=J \sum \limits_{n} [ (1+\delta){\bf \hat S}_{1,n} \cdot {\bf \hat S}_{2,n} 
+ (1-\delta){\bf \hat S}_{2,n} \cdot {\bf \hat S}_{1,n+1}] 
\end{equation}
\noindent where the dimerization parameter $\delta$ lies in the range
$[0,1]$. We find that the ground state and low-energy excitations are
qualitatively similar to the undimerized case, $\delta=0$. Namely, the
ground state has $s_{tot.}=N(s_{1}-s_{2})$. There is a gapless branch
of excitations $\omega_{1k}$ with $s_{tot.}=N(s_{1}-s_{2}) -1$ and dispersion
\begin{equation}
\omega_{1k}= J(-s_{1}+s_{2}) + J\sqrt{(s_{1}-s_{2})^{2}+ 4s_{1}s_{2}
(1-\delta^{2}) \sin^{2}(k/2)} ,
\end{equation}
\noindent and a gapped branch of excitations $\omega_{2k}$ with 
$s_{tot.}=N(s_{1}-s_{2}) +1$ and whose dispersion is, 
\begin{equation}
\omega_{2k}=\omega_{1k}+2J(s_{1}-s_{2})
\end{equation}
To this order in spin wave theory, the gap at $k=0$ is given by
$\Delta=2J(s_{1}-s_{2})$ independent of $\delta$. Numerically, however
we will see below that the gap increases almost linearly with $\delta$ 
for the spin-1 / spin-{$\frac{1}{2}$} alternating spin chain.

\subsection{Variational Calculation}

The short correlation length found above for the alternating 
spin-1 / spin-{$\frac{1}{2}$} chain
suggests that a $product$ wave function\cite{aklt} could
be a good variational trial state for obtaining the ground state
properties. We work in the Fock space basis of the alternating spin
chain. We are interested in a variational wave function in which
the state $|\ldots~~~1~~~~ -\frac{1}{2}~~~~1~~~~ -\frac{1}{2}~~~ \ldots>$ 
of fig.1  is
expected to have the largest amplitude, say, 1. We can then show that
a state in which there are $n$ spins on the spin-1 sublattice in the state
 $s^{z}=0$ must have a real amplitude with the sign $(-1)^{n}$. 
This sign rule
(which is analogous to the Marshall sign rule for the spin-{$\frac{1}{2}$} 
chain\cite{marshall})
can be proved by using the Perron-Frobenius theorem and the fact that
the states with an odd number of spins in the $s^{z}=0$ state 
are connected to states
with an even number of spins in the $s^{z}=0$ states 
by the operation of an odd number of exchange terms in the Hamiltonian (Eq.1).

Let us now introduce the following real and positive amplitudes for the
six states possible for a nearest neighbour bond, namely, unit amplitude for
$|1, -\frac{1}{2}>$ and $|-1, \frac{1}{2}>$, an amplitude $\eta_{1}$ for 
$|0, -\frac{1}{2}>$
 and $|0, \frac{1}{2}>$ and amplitude $\eta_{2}$ for $|-1, -\frac{1}{2}>$ and 
$|1, \frac{1}{2}>$.
Now, we consider a variational wave function of the form 
\begin{equation}
|\Psi(\eta_{1},\eta_{2})> = \sum \limits_{a} C_{a}|\psi_{a}>
\end{equation}
\noindent where $a$ runs over all the $6^{N}$ possible states of N
 spin-1 and spin-{$\frac{1}{2}$} sites and $C_{a}=(-1)^{n} \prod \limits_{j} 
\omega_{a,j}$,
 where the product is over all the $2N$ bonds, $\omega_{a,j}$ is the
 amplitude
of the bond $j$ and $n$ denotes the number of spins in the 
$s^{z}=0$ state in the chain state $a$.

The state $|\Psi(\eta_{1}, \eta_{2})>$ is translationally invariant. 
However, it is not an eigenstate of the total spin ${\hat S}^{2}_{tot.}$ or 
even ${\hat S}^{z}_{tot.}$; 
further, since
it includes states $\{s^{z}_{n}\}$ and $\{-s^{z}_{n}\}$ with equal 
amplitude, the expectation value of ${\bf \hat S}^{z}_{1,n}$ or
${\bf \hat S}^{z}_{2,n}$
is zero for any site $n$. In spite of these drawbacks, we will see below
that the $|\Psi(\eta_{1}, \eta_{2})>$ gives a good variational ground 
state energy for an appropriate choice of $\eta_{1}$ and $\eta_{2}$. 

Calculations involving the state $|\Psi(\eta_{1}, \eta_{2})>$ can be 
carried out using
the transfer matrix method. For instance, the probabilities of the six
possible bonds for $|\frac{1}{2}, n> \bigotimes <1, n+1|$ are given by the
$2 \times 3$ matrix. \\
$$A = \pmatrix{ \eta_{2}^{2} & \eta_{1}^{2} & 1 \cr
                 1 & \eta_{1}^{2} & \eta_{2}^{2} } $$
\noindent while the probabilities of the bonds $|1,n> \bigotimes 
<\frac{1}{2},n|$ are
given by the $3 \times 2$ matrix $A^{T}$. Then the norm 
$<\Psi(\eta_{1}, \eta_{2})| \Psi(\eta_{1}, \eta_{2})> = Tr (AA^{T})^{N}$. 
The matrix $AA^{T}$
has the eigenvalues $\lambda_{+}=(1+\eta_{2}^{2})^{2} + 2\eta_{1}^{4}$ and
$\lambda_{-}= (1-\eta_{2}^{2})^{2}$. Hence, $<\Psi(\eta_{1}, \eta_{2})|
\Psi(\eta_{1}, \eta_{2})> =\lambda^{N}_{+} + \lambda^{N}_{-}$, 
which is dominated by $\lambda^{N}_{+}$ as $N \rightarrow \infty$. 

We can now calculate the variational energy per bond $\epsilon(\eta_{1}, 
\eta_{2})$
through the expression
\begin{equation}
\epsilon(\eta_{1}, \eta_{2}) <\Psi(\eta_{1}, \eta_{2})| \Psi(\eta_{1},
\eta_{2})>
 =\frac{1}{2N} <\Psi(\eta_{1}, \eta_{2})| H |\Psi(\eta_{1}, \eta_{2})>
\end{equation}
This is equal to $<\Psi(\eta_{1}, \eta_{2})| {\bf \hat S}_{\frac{1}{2},n} 
\cdot 
{\bf \hat S}_{1,n+1}|\Psi(\eta_{1}, \eta_{2})>$
 by translational invariance. We find that, \\
$\lefteqn{\epsilon(\eta_{1}, \eta_{2})=}$
\begin{eqnarray}
J\frac{[-\frac{1}{2}(1+\eta_{2}^{2})^{3}(1-\eta_{2}^{2}) - 2\sqrt{2}
(1+\eta_{2}^{2}) \eta_{1}^{2} \eta_{2} (1+\eta_{2}) -\eta_{1}^{4}(1-\eta_{2}^
{4}) -2\sqrt{2}
\eta_{1}^{6}(1+\eta_{2})]}{[(1+\eta_{2}^{2})^{2}+ 2\eta_{1}^{4}]^{2}}
\end{eqnarray}
We now minimize this function of $\eta_{1}$ and $\eta_{2}$, and 
 find
that the minimum occurs at $\eta_{1}=0.842$, $\eta_{2}=0.445$, 
 giving $\epsilon=-0.701J$.
This compares favourably with the spin wave result of $-0.718J$.

We can compute the two-spin correlation function and determine how it 
decays asympotically at large distances. The correlation length $\xi$
for the asymptotic decay is given by, $\xi^{-1}=\ln(\lambda_{+}/\lambda_{-})$.
Hence $\xi$ is $0.749$ at the values of $\eta_{1}$, $\eta_{2}$ given above. 
This is even shorter
than the value of $1.44$ found from the spin wave theory.

We can improve the variational calculation by allowing the five 
independent ( taking the amplitude for $|1, -\frac{1}{2}>=1$) real 
amplitudes for 
 the different possible states of the 
nearest neighbour bonds  rather
than only two amplitudes ( $\eta_{1}$, $\eta_{2}$) as used above. However 
we will not do so here as such a calculation would be tedious and the
two-amplitude {\it ansatz} has already given us a good understanding of the 
short correlation length and the ground state energy.

\subsection{DMRG Studies}

 We have performed DMRG calculations on alternating 
spin-1 / spin-{$\frac{1}{2}$}
 chain
with open boundary condition for the Hamiltonian in Eq.(1). We compute
the ground state properties by studying chains of upto 50 to 100 sites.
The number of dominant density matrix eigenstates,
 $m$, that we have retained at each DMRG iteration is between 80 to 100. 
The DMRG procedure follows the usual steps
for chains discussed in earlier papers\cite{white1,dmrg,skp} except that 
the chains do not
have the symmetry between the left and right halves and the density
matrices for these two halves have to be constructed at every iteration
of the calculations. The DMRG results reported in this paper
are all specialized to the case of $s_{1}=1$ and $s_{2}=\frac{1}{2}$.
The ground state of the chain lies in the $M_{s}=N(s_{1}-s_{2})$ sector
for a $2N$ sites system and we cannot use spin parity symmetry in this sector. 
In general, for $M_{s} \ne 0$, parity cannot be exploited, keeping 
$M_{s}$ a good quantum number.
The ground state is confirmed to be in $M_{s}=N(s_{1}-s_{2})$ 
sector for a chain of $2N$ sites from extensive checks carried out by
obtaining the low-energy eigenstates in different $M_{s}$ sectors of a 
20-site chain. A state corresponding to the lowest energy in 
$M_{s}=N(s_{1}-s_{2})$ is found in all subspaces with 
$|M_{s}| \le N(s_{1}-s_{2})$ and is absent in subspaces with 
$M_{s} > N(s_{1}-s_{2})$. This shows that the spin in the ground state,
 $s_{G}=N(s_{1}-s_{2})$.
In fig.3, we show the extrapolation of the energy per site
as function of inverse system size. The energy/site, $\epsilon_{0}$, 
extrapolates to  
$-0.72704J$ for the ground state of the system. 
The spin wave analysis value of $-0.718J$ for $\epsilon_{0}$ 
is higher than
the DMRG value. It is worth mentioning at this point that the DMRG
ground state energy/site agrees to better than $10^{-7}$ with the
 exact solution in the case of uniform spin-{$\frac{1}{2}$} Heisenberg
 antiferromagnet\cite{white1,white2}. 
It is interesting to note that, in the alternating spin case, the 
energy/site lies in between the values for the pure spin-{$\frac{1}{2}$} 
 uniform chain ( $-0.443147J$ ) \cite{pear} and the pure spin-1 uniform 
chain ( $-1.401484J$ ) \cite{white2}.

In fig.4, we show the spin-density at all the sites
of a chain of 100 sites. The spin-density is uniform on each of the 
sublattices in the chain.
The expectation value of ${\bf \hat S}_{z}$ at the spin-1 site reduces 
from the classical
value of $1$ to $0.79248$, while, at the spin-{$\frac{1}{2}$} site, 
it is $-0.29248$.
This can be compared with spin wave values of $0.695$ and $-0.195$ for
spin-1 and spin-{$\frac{1}{2}$} sites respectively. We note that the
spin wave analysis overestimates the quantum fluctuations.
 It is very interesting to note that, the spin-density
distribution in alternating spin chain behaves more like in a ferromagnetic 
chain than like in an antiferromagnet,
 with the net spin of each unit cell perfectly aligned ( but with small
fluctuations on the individual sublattices).
In a ferromagnetic ground state the spin-density at each site has 
the classical value appropriate
to the site spin, whereas for an antiferromagnet, this 
averages out to zero at each site as the ground state is nonmagnetic.
 From this viewpoint, the ferrimagnet is similar to a ferromagnet and
 is quite unlike an antiferromagnet. The spin wave analysis also yields the
same physical picture.

 Owing to the alternation 
of the
spin-1 and spin-{$\frac{1}{2}$} sites along the chain, one has to 
distinguish between $<{\bf \hat S}^{z}_{1,0} {\bf \hat S}^{z}_{1,n}>$, 
$<{\bf \hat S}^{z}_{2,0} {\bf \hat S}^{z}_{2,n}>$ and 
$<{\bf \hat S}^{z}_{1,0} {\bf \hat S}^{z}_{2,n}>$ pair correlations.
We calculate all the three correlation functions with the mean values
subtracted out as in Eq.(11), since the mean values are nonzero in this
system unlike in a pure antiferromagnetic spin chain. In the DMRG procedure,
 we have computed these correlation functions from the sites inserted at
the last iteration, to minimize numerical errors. In fig.5, 
we plot the two-spin correlation functions in the ground state as a 
function of the distance
between the spins for an open chain of 100 sites. 
All three correlation functions decay rapidly with distance. 
From the figure it is clear that, except
for $<{\bf \hat S}^{z}_{1,0} {\bf \hat S}^{z}_{2,n}>$ correlation, 
other correlations are 
almost zero even for the shortest possible distances. 
The $<{\bf \hat S}^{z}_{1,0} {\bf \hat S}^{z}_{2,n}>$ correlation
has a finite value ( $-0.094$ ) only for the nearest neighbours. 
This rapid decay of the correlation functions do not easily allow one
to find the exact correlation length $\xi$ for a lattice model 
though it is clear that $\xi$ is very small
($i.e.$ less than one unit). As mentioned earlier, spin wave theory gives
$\xi =1.44$ while the variational calculation gives $\xi=0.75$.
This type of decay could
be compared with the behaviour of the correlation function
 at the Majumdar-Ghosh
point\cite{majumdar} for the pure spin-{$\frac{1}{2}$} Heisenberg chain
with nearest and next nearest neighbour antiferromagnetic exchange 
interactions ( $J_{1}-J_{2}$ model at $J_{2}=0.5J_{1}$). The AKLT model
at the exactly solvable point\cite{aklt} in the case of a pure spin-1 
chain described by a bilinear-biquadratic Hamiltonian also has a very short
correlation length ( $\xi=0.91$ ).
Both the cases compared above, however, have one property in
common, $i.e.$, the exactly solvable point in both models lie in
the gapped phase.  

The lowest spin excitation of the alternating spin chain is to a state 
with $s=s_{G}-1$. 
To get this state, we target the 2nd state in $M_{s}=s_{G}-1$ sector 
of the chain. To confirm that this state is the $s=s_{G}-1$
state, we have computed the 2nd state in $M_{s}=0$ sector and find that
it also has the same energy. However, the corresponding state is
absent in $M_{s}$ sectors with $M_{s} > | s_{G}-1 |$. Besides, from exact 
diagonalization of all the states of an alternating spin
chain with 8 sites, we find that the ordering of the state is such that 
the lowest excitation is to the state with spin $s=s_{G}-1$. 
We have obtained the excitation gap of the alternating spin chain in the
limit of infinite chain length by extrapolating from the plot of spin gap
$vs$ the inverse of the chain length ( fig.6 ). 
We find that this excitation is gapless in the infinite chain limit.
This is very unusual in that the correlation function in the ground state
decays exponentially ( in fact $\xi < 1$) but the system
has a gapless excitation, in agreement with the spin wave analysis.

Motivated by the spin-wave
calculation, we also have computed the energy of the $s=s_{G}+1$ state
by targetting the lowest state in $M_s=s_{G}+1$ sector. In fig.7,
we plot the excitation gap to the $s=s_{G}+1$ state from the ground state
as a function of the inverse of the chain length. The gap saturates 
to a finite value of $(1.2795 \pm 0.0001)J$. The $S_z$ expectation values 
computed in this state are found to be uniform on each of the
sublattices ( independent of the site)
and leading us to believe that this excitation is not a magnon like 
excitation ( quantized in a box ) as has been observed for a 
spin-1 chain in the Haldane phase\cite{white2}. 

It would be interesting to know the total spin of the states as a 
function of their energies. For a smaller alternating spin system,
 it is possible to characterize all the states by their energy and
total spin value, by resorting to an exact diagonalization scheme.
The total spin value of a state is naturally fixed if we exploit
the total spin conservation property of the Hamiltonian while
constructing the Hamiltonian matrix. This can, for example, 
achieved by using a valence bond basis\cite{srsoos} for setting up
the Hamiltonian matrix. Alternately, we can also compute the
expectation value of the total spin operator in each eigenstate
in the $M_{s}=0$ sector to provide a spin label for each of the
states. We have followed the latter procedure. In fig.8, we present
the energy levels as a function of the total spin of the states for an
eight site ferrimagnetic ring and twelve site spin-{$\frac{1}{2}$} 
ferromagnetic and antiferromagnetic rings.
We find that the spin of the state appears to vary irregularly with energy
unlike in the case of pure spin-{$\frac{1}{2}$} ferro and antiferromagnets
(fig.8). Careful comparison of the three plots in the figure shows that
the low-lying excitations of the ferrimagnet to spin states $s_{tot} < s_{G}$
is ferromagnetic like and to states with $s_{tot} > s_{G}$ is
antiferromagnetic like, for finite systems.

We have also studied the dimerized alternating spin chain defined by the
Hamiltonian ( Eq.12 ) with $\delta$, the dimerization parameter, in the range
$0<\delta \le 1$.
Earlier works on spin chains\cite{sumit} have revealed that with the 
alternation $\delta$ in the exchange parameter, the  half integer spin 
chain will have an unconditional
spin-Peierls transition whereas for integer spin chain the transition is
conditional. This conclusion has been drawn from the fact that, with
the inclusion of $\delta$, the magnetic energy gain $\Delta E$  
can be defined as,
\begin{equation}
    \Delta E(2N,\delta) = {1 \over 2N} [ E(2N,\delta) - E(2N,0) ]
\end{equation}
\noindent where $E(2N,\delta)$ is the ground state energy of the 
$2N$ sites system 
with alternation $\delta$ in the exchange integral and $E(2N,0)$ is the 
ground state energy of the uniform chain of $2N$ sites. For the pure 
spin chain, if we
assume that $\Delta E$ varies as $\delta^{\nu}$ for small $\delta$,
we find that $\nu=4/3$ for the spin-{$\frac{1}{2}$} chain 
and $\nu=2$ for the spin-1 chain\cite{sumit}.
Thus, for the spin-{$\frac{1}{2}$}
chain, the stabilization energy always overcomes the elastic energy,
whereas for the spin-1 case, it depends on the stiffness of the lattice.

We have employed a DMRG calculations to obtain $\Delta E(2N,\delta)$, for
small values of $\delta$ for the alternating spin chain.
To determine the exact
functional form of the magnetic energy gain, we varied the chain length
from 50 sites to 100 sites and also $m$( the number of states retained
in each
DMRG iteration) from 80 to 100 to check the convergence of
$\Delta E$ with the chain length. The dependence of $\Delta E(2N,\delta)$ on
$1/2N$ is linear for the $\delta$ values we have studied. Fig.9
gives a sample variation of $\Delta E(2N, \delta)$ upon $1/2N$. This
allows us to extrapolate $\Delta E(2N,\delta)$ to the infinite chain
limit reliably. In fig.10, we show the plot of $\Delta E(2N,\delta)$ vs. 
$\delta$ for finite $2N$ values and also the extrapolated infinite chain.
We see that there is a gain in magnetic energy upon dimerization even in the
infinite chain limit. To obtain the exponent $\nu$, we plot 
$\ln \Delta E(2N,\delta)$ vs. $\ln \delta$ for the infinite
chain ( fig.11 ). From this figure, we find that
in the alternating spin case, for the infinite chain 
$\Delta E \approx \delta^{2.00 \pm 0.01}$.
Thus, the spin-Peierls transition appears to be close to being conditional 
in this system. The magnetic energy gain/site for finite chains is larger
than that of the infinite chain for any $\delta$ values ( fig.10). 
It is possible that the distortion in finite chain is unconditional while
that of the infinite chain is conditional.

We have also studied the spin excitations in the dimerized alternating
spin-1 / spin-{$\frac{1}{2}$} chain.
We calculate the lowest spin excitation to the $s=s_{G}-1$ state
from the ground state. We find that the $s=s_{G}-1$ state is gapless from 
the ground state for all $\delta$ values. This result agrees with the
spin wave analysis. The system remains gapless even while
dimerized unlike the pure antiferromagnetic dimerized spin chains. 
There is a smooth 
increase of the spin excitation gap to $s=s_{G}+1$ state from ground state 
with increasing $\delta$. We have plotted this gap with $\delta$ 
in fig.12.  The gap shows almost a linear behaviour
as a function of $\delta$, with an exponent of $1.07 \pm 0.01$.
The spin wave analysis however shows that this excitation gap is independent
of $\delta$.

\section { Low-Temperature Properties }

In this section, we present results of our DMRG calculations of the
thermodynamic properties of the alternating spin system. The size of the 
system varies from 8 to 20 sites. We impose 
periodic boundary conditions to minimize finite size effects.
We set-up the Hamiltonian matrices in the DMRG basis for all allowed 
$M_{s}$ sectors for a ring of 2N sites. We can diagonalize these matrices
completely to obtain all the eigenvalues in each of the $M_{s}$ sectors. 
As the number of DMRG basis states increases rapidly with 
increasing m, we retain a smaller number of dominant density matrix
eigenvectors in the DMRG procedure, $i.e.$, $50 \le m \le 65$, 
depending on the $M_{s}$ sector as well as the size of the system.
We have checked to find the dependence of properties ( with $m$ in the
range $50 \le m \le 65$ ) for the system sizes we have studied 
( $8 \le 2N \le 20$ ) and have confirmed that the properties do not
vary significantly for the temperatures at which they are computed.
The above extension of the DMRG procedure is found to be accurate
by comparing with exact diagonalization results for small systems.
It may appear surprising that the DMRG technique which essentially
targets a single state, usually the lowest energy state in a chosen
sector, should provide accurate thermodynamic properties since these
properties are governed by energy level spacings and not the absolute
energy of the ground state. However, there are two reasons why the
DMRG procedure yields reasonable thermodynamic properties. Firstly,
 the projection of the low-lying excited state eigenfunctions on the
DMRG space in which the ground state is obtained is substantial
and hence these excited states are well described in the chosen
DMRG space. The second reason is that the low-lying excitations
of the full system are often lowest energy states in different
sectors in DMRG procedure and thus their energies are quite accurate 
even on an absolute scale.

The canonical partition function $Z$ for the 2N site ring can be written as
\begin{equation}
Z= \sum_{j}{e^{-\beta ( E_{j} - B(M_{s})_{j} )}}
\end{equation}
\noindent where, the sum is over all the DMRG energy levels of the
 2N site system in all the $M_{s}$ sectors.
 $E_{j}$ and $(M_{s})_{j}$ are energy and z-component of the total spin
 of the state $j$, B is the strength of the magnetic field in units of
$J/g\mu_{B}$ ( $g$ is the gyromagnetic ratio and $\mu_{B}$ is the
Bohr magneton ) along ${\bf z}$ direction and $\beta=J/k_{B}T$ with 
$k_{B}$ and T being the Boltzmann constant and temperature respectively.
The field induced magnetization, $< M >$,  can be defined as
\begin{equation}
< M >=\frac{\sum_{j}(M_{s})_{j}{e^{-\beta ( E_{j} - B(M_{s})_{j} )}}}{Z}
\end{equation}
\noindent the magnetic susceptibility, $\chi$, by relating it to the 
fluctuation in magnetization,
\begin{equation}
\chi=\beta [ < M^{2} > - < M >^{2} ]
\end{equation}
\noindent and similarly the specific heat, $C$, by relating it to the 
fluctuation in energy, can be written as,
\begin{equation}
C = \frac{\beta}{T} [ < E^{2} > - < E >^{2} ]
\end{equation}

The dimensionalities of the DMRG Hamiltonian matrices that we completely 
diagonalize
vary from $2500$ to $3000$, depending upon the DMRG parameter $m$ and
the $M_s$ value of the targetted sector, for rings of sizes greater
than $12$. These matrices are not very sparse, owing to the cyclic
boundary condition imposed on the system. The DMRG properties  compare
very well with exact results for
small system sizes amenable to exact diagonalization studies. In the
discussion to follow, we present results on the 20-site ring although
all calculations have been carried out for system sizes from 8 to 20 sites.
This is because the qualitative behaviour of the properties we have studied
are similar for all the ring sizes in this range.

We present the dependence of magnetization on temperature for different
magnetic field strengths in fig.13. At low magnetic fields, the 
magnetization shows a sharp decrease at low-temperatures and 
shows paramagnetic behaviour at high temperatures. As the field strength
is increased, the magnetization shows a slower decrease with temperature
and for field strength comparable to the exchange constant, the 
magnetization shows a broad maximum. This behaviour could be understood 
from the type of spin excitations present in the system. The lowest
energy excitation at low magnetic fields is to a state with spin $s$
less than $s_G$. Therefore, the magnetization initially decreases at
low temperatures. As the field strength is increased, the gap to 
spin states with $s > s_G$ decreases as the Zeeman coupling to these
states is stronger than to the states with $s \le s_G$. The behaviour
of the system at even stronger fields turns out to be remarkable.
The magnetization in the ground state ( $T=0$ ) shows an abrupt increase
signalling that the ground state at this field strength has $M_{s} >s_G$.
The temperature dependence of the magnetization shows a broad 
maximum indicating the presence of states with even higher spin values
lying above the ground state in the presence of this strong field.
Only at very intense fields do we find the magnetization decreasing
with increasing temperature. This happens because at such large
field strengths, the ground state is the highest spin state possible
for the system. 

The dependence of magnetization on the magnetic field is shown at 
different temperatures in fig.14. At low temperature the magnetization
shows a plateau. The width of the plateau decreases as the temperature
is raised and eventually the plateau disappears. The existence of
the plateau shows that the higher spin states are not accessible 
at the chosen temperature. At higher fields, the larger Zeeman
splittings of higher spin states become accessible leading to an
increase in the magnetization. All these curves intersect at 
$B=J/g\mu_{B}$ and $B=2.5 J/g\mu_{B}$ and these fields are close to 
the field strengths at which the ground state switches from one 
$M_s$ value to another higher value.

The dependence of ${\chi T}/2N$ on temperature for different
field strengths are shown in fig.15. For zero field, the zero temperature
value of $\chi T$
is infinite in the thermodynamic limit and for finite rings is 
finite and equal to the fluctuation in magnetization. For the ferrimagnetic
ground state ${\chi T}/2N$, as $T \rightarrow 0$, is given by
$s_{G}(s_{G}+1)/6N$. As the 
temperature increases, this product decreases and shows a minimum
around $k_BT=0.5J$ before increasing again. The minimum manifests
due to the states with $M_s < s_G$ getting populated at 
low-temperatures. These states in the infinite chain limit turn out
to be the gapless excitations in the spectrum.  The subsequent
increase in the $\chi T$ product is due to the higher energy and
higher spin states accessed with further increase in temperature.
Experimentally, it has been found in the bimetallic chain compounds
that the temperature at which the minimum occurs in the $\chi T$ 
product depends upon the magnitude of the spins $s_1$ and $s_2$\cite{kahn3}.
The $Ni^{II}-Cu^{II}$ bimetallic chain shows a minimum
in ${\chi T}/2N$ at a temperature corresponding to $55$ $cm^{-1}$ ($80 K$)
and independent estimate of the exchange constant in this system is
$100$ $cm^{-1}$\cite{kahnpc}. This is in very good agreement with the
minimum theoritically found at temperature $(0.5 \pm 0.1)J$. The minimum in 
${\chi T}/2N$
vanishes at $B=0.1 J/g\mu_{B}$ corresponding to $\sim 10 T$ and it
would be interesting to study the magnetic susceptibility of other
systems experimentally under such high fields. The low-temperature 
zero-field behaviour of our system
can be compared with the one-dimensional ferromagnet. In the
latter, the spin wave analysis shows that the $\chi T$ 
product increases as $1 \over T$ at low temperatures\cite{taka87}. 

In finite but weak field, the behaviour of $\chi T$ is different. 
The magnetic field opens up a gap and $\chi T$ goes exponentially 
to zero for temperatures less than the gap in the applied field. 
Even in this case a minimum is found at the same temperature as in
the zero-field case, for the same reason discussed in the zero field
case.

In stronger magnetic fields, the behaviour of $\chi T$ 
from zero temperature upto $k_BT=0.5J$ is qualitatively different. 
The minimum in this case vanishes. In these field strengths, the
states with higher $M_s$ values are accessed even below $k_BT=0.5J$.
The dependence of $\chi T$ above $k_BT=0.5J$ is the same in all cases.
In even stronger magnetic fields, the initial sharp increase is
suppressed. At very low temperature, the $\chi T$ product is nearly
zero and increases linearly with $T$ over the temperature range
we have studied. This can be attributed to a switch in the ground state
at this field strength. The very high temperature behaviour of $\chi T$
should be independent of field strength and should saturate to the
Curie law value corresponding to the mean of spin-1 and spin-{$\frac{1}{2}$}
values which is $11/24$.

The temperature dependence of specific heat also shows marked
dependence on the magnetic field at strong-fields. This dependence
is shown in fig.16 for various field strengths. In zero and weak
magnetic fields, the specific heat shows a broad maximum around
$k_BT=0.6J$. At strong-magnetic field ($B=J$), there is a dramamtic
increase in the peak height at about the same temperature, although
the qualitative dependence is still the same as at low-magnetic fields.
This indicates that the higher energy high-spin states are brought
to within $k_BT$ of the ground state at this magnetic field strength. 

Studies on dimerized alternating spin chains show qualitatively
similar trends as the uniform chains. This is not surprising as
the low-energy spectrum of the system does not change qualitatively
upon dimerization.

\section{Summary}

We have studied the alternating spin-1 / spin-{$\frac{1}{2}$} model
in detail. The ground and low-lying excited states have been
analyzed by using a spin wave theory as well as DMRG calculations.
Both the methods predict a ground state with spin $s_G=N(s_1-s_2)$
for a $2N$ site system. They also predict a gapless excitation to 
a state with $s=s_G-1$ in the infinite chain limit. The lowest
gapped excitations are to states with spin $s=s_G+1$. The very
short correlation length in the ground state of the system 
motivated its description by a variational
trial function of the product type. Interestingly, the spectrum 
is qualitatively unchanged upon dimerization. The dimerization is 
itself conditional in the infinite chain limit.

The DMRG technique is also employed to obtain the low-temperature
thermodynamic properties. The magnetic susceptibility shows very
interesting magnetic field dependence. The $\chi T$ vs. $T$ plot
shows a minimum at low-magnetic fields. This minimum vanishes at 
high-magnetic fields. The specific heat shows a maximum as a 
function of temperature at all fields. The height of the maximum 
shows a dramatic increase at high-magnetic field. Experimental
systems describable by this model exist and have been studied
quite extensively. It is hoped that our studies will motivate 
experimental studies of these systems in high magnetic fields.

{\bf Acknowledgements}
We are thankful to Professor Olivier Kahn, who by introducing one of us (S.R.) 
to the experimental alternating spin systems, motivated us to undertake
this work. The authors are grateful to Dr. R. Chitra
and Ms. Y. Anusooya for many helpful discussions. The
present work has been supported by the Indo-French
Centre for the Promotion of Advanced Research through
project No. 1308-4.

\pagebreak

{\centerline{\bf Figure Captions}} 

\noindent{\bf Fig.1} \\
 Schematic picture of the arrangement of spins $s_{1}$ and $s_{2}$ on a
chain with interactions discussed by the Hamiltonian in
Eq.(1). 

\noindent{\bf Fig.2} \\
The two branches of the spin wave dispersion curves, $\omega_{1k}$ and
$\omega_{2k}$ ( Eq.(7) ), in units of $J$, for the alternating spin system
with $s_{1}=1$ and $s_{2}=\frac{1}{2}$.

\noindent{\bf Fig.3} \\
Extrapolation of the ground state energy/site ($\epsilon_{0}$), in units of
$J$, as a function of inverse
system size, for the uniform alternating spin-1 /spin-{$\frac{1}{2}$} chain.

\noindent{\bf Fig.4} \\
Expectation value of the $z$-component of the site spin vs. site index, $n$.
The upper and lower points are for the spin-1 ($i=1$) and the 
spin-{$\frac{1}{2}$} ($i=2$) sites respectively.

\noindent{\bf Fig.5} \\
Subtracted two-spin correlation functions (defined in the text)
as a function of distance between the two spins.
(a) spin-1 spin-1 correlations, (b) spin-{$\frac{1}{2}$} 
spin-{$\frac{1}{2}$} correlations and (c) spin-1 spin-{$\frac{1}{2}$} 
correlations.

\noindent{\bf Fig.6} \\
Energy gap (units of $J$) from the ground state to the lowest state with 
spin $s=s_{G}-1$ as a function of inverse system size. $s_{G}$ is the 
total spin of the ground state.

\noindent{\bf Fig.7} \\
Excitation gap (units of $J$) from the ground state (spin $s=s_{G}$) to the
state with spin $s=s_{G}+1$ as a function of the inverse system size. 
The convergence to the infinite system is much faster for this gapped 
excitation, as compared to the gapless excitation in fig.6.

\noindent{\bf Fig.8} \\
Plot of energy (units of $J$) vs. the total spin quantum number of the
complete
spectrum of (a) an 8 site ring with an alternating spin-1 / 
spin-{$\frac{1}{2}$}
arrangement (b) a ring of 12 site spin-{$\frac{1}{2}$} antiferromagnet and 
(c) a ring of 12 site spin-{$\frac{1}{2}$} ferromagnet. The states with 
very high energies are not shown, for (b) and (c).

\noindent{\bf Fig.9} \\
Gain in magnetic energy (units of $J$) associated with dimerization vs. 
the inverse system size for three different values of dimerization $\delta$. 
(a) $\delta=0.025$, (b) $\delta=0.05$ and (c) $\delta=0.075$.

\noindent{\bf Fig.10} \\
Magnetic energy gain $\Delta E(2N, \delta)$ (units of $J$) as a function 
of dimerization parameter $\delta$ for different system sizes. In the figure 
$2N=50$ (squares), $2N=100$ (circles) and extrapolated values with
$N \rightarrow \infty$ (triangles) are shown.

\noindent{\bf Fig.11} \\
Log-log plot of extrapolated magnetic energy gain (units of $J$) for infinite 
system size and dimerization parameter $\delta$. The slope is calculated
to be $2.00 \pm 0.01$.

\noindent{\bf Fig.12} \\
Excitation gap (units of $J$) to the state with spin $s=s_{G}+1$ from the 
ground state ($s=s_{G}$)  as a function of $\delta$ for the dimerized 
alternating spin-1 / spin-{$\frac{1}{2}$} chain.

\noindent{\bf Fig.13} \\
Plot of magnetization/site as a function of temperature $T$ for five 
different values of magnetic fields, $B$. Squares are for $B=0.1 J/g\mu_{B}$, 
circles for $B=0.5 J/g\mu_{B}$, triangles for $B=J/g\mu_{B}$, diamonds for 
$B=2 J/g\mu_{B}$ and inverse triangles for $B=3 J/g\mu_{B}$.

\noindent{\bf Fig.14} \\
Magnetization/site vs. the magnetic field strength $B$, in units of 
$J/g\mu_{B}$, for four different temperatures $T$. $T=0.3 J/k_{B}$
results are
given by squares, $T=0.5 J/k_{B}$ by circles,
$T=0.7 J/k_{B}$ by triangles and $T=J/k_{B}$ by diamonds.

\noindent{\bf Fig.15} \\
$\chi T$ ( defined in the text) per site as a function of temperature $T$
for various magnetic field strengths, $B$. Zero field results are
shown by squares, $B=0.01 J/g\mu_{B}$ by circles, $B=0.1 J/g\mu_{B}$ by 
triangles and $B=J/g\mu_{B}$ by diamonds.

\noindent{\bf Fig.16} \\
Specific heat/site as a function of temperature $T$ for four different
values of magnetic fields, $B$. Zero field data are shown by squares,
$B=0.01 J/g\mu_{B}$ by circles, $B=0.1 J/g\mu_{B}$ by triangles and 
$B=J/g\mu_{B}$ by diamonds.

\end{document}